\documentclass[aps,prc,groupedaddress,showpacs,manuscript]{revtex4}

\usepackage{graphicx}

\begin{document}

\title{$\Sigma^-/\Sigma^+$ ratio as a candidate for probing the density dependence of the symmetry
potential at high nuclear densities }

\author {Qingfeng LI$^{1,4)}$\email[]{Qi.Li@fias.uni-frankfurt.de}, Zhuxia LI$^{2,1,3,4)}$\email[]{lizwux@iris.ciae.ac.cn}, Enguang ZHAO$^{4)}$ and Raj K. GUPTA$^{1,5)}$ }
\address{
1) Frankfurt Institute for Advanced Studies (FIAS), Johann Wolfgang Goethe-Universit\"{a}t, Max-von-Laue-Str. 1, D-60438 Frankfurt am Main, Germany\\
2) China Institute of Atomic Energy, P. O. Box 275 (18),
Beijing 102413, P. R. China\\
3) Center of Theoretical Nuclear Physics, National Laboratory of
Lanzhou Heavy Ion Accelerator,
 Lanzhou 730000, P. R. China\\
4) Institute of Theoretical Physics,
Chinese Academy of Sciences, P. O. Box 2735, Beijing 100080, P. R. China\\
5) Department of Physics, Panjab University, Chandigarh - 160014, India
 }


\begin{abstract}
   Based on the UrQMD model, we have investigated the influence of the
symmetry potential on the negatively and positively
charged $\pi$ and  $\Sigma$ hyperon production ratios in heavy ion collisions at the
SIS energies. We find that, in addition to $\pi^-/\pi^+$ ratio, the $\Sigma^-/\Sigma^+$ ratio can
be taken as a sensitive probe for investigating the density
dependence of the symmetry potential of nuclear matter at high
densities (1-4 times of normal baryon density). This sensitivity of the symmetry
potential to both the $\pi^-/\pi^+$ and $\Sigma^-/\Sigma^+$
ratios is found to depend strongly on the incident beam energy. Furthermore, the $\Sigma^-/\Sigma^+$ ratio is shown to carry the information about
the isospin-dependent part of the $\Sigma$ hyperon single-particle potential.
\end{abstract}


\pacs{24.10.Lx, 25.75.Dw, 25.75.-q} \maketitle

\section{Introduction}
The equation of
state (EoS) has attracted a lot of attention recently for asymmetric nuclear matter, which can be described approximately
by the parabolic law
\begin{equation}
e(\rho,\delta)=e_{0}(\rho,0)+e_{sym}(\rho)\delta^{2}. \label{ieos}
\end{equation}
Here $\delta=(\rho_{n}-\rho_{p})/(\rho_{n}+\rho_{p})$ is the
isospin asymmetry, $e_{0}$ the energy per nucleon for symmetric
nuclear matter, and $e_{sym}(\rho)$ the bulk symmetry energy. The
symmetry energy term $e_{sym}(\rho)\delta^{2}$ is very important
for understanding many interesting astrophysical phenomena (see,
{\it e.g.} \cite{Li98}), but so far results in large
uncertainties: {\it e.g.}, the symmetry energy calculated with
different kinds of parameter sets (Skyrme or Gogny type) are
largely divergent \cite{Brown2000,Margueron2001}), especially at
high densities, and for some cases, {\it i.e.}, when the density
is higher than three times of the normal density, even a negative
symmetry energy can be obtained. Therefore, acquiring the more
accurate knowledge of the symmetry energy, and the isospin
asymmetry, becomes one of the main goals in nuclear physics at
present. The recently available facilities of rare-isotope beams
provide the opportunities to study the dynamical evolution of
nuclear systems with a large range of isospin asymmetries, which
increases the domain over which a spatially uniform local isospin
asymmetry $\delta(r)$ may be achieved.

  In order to obtain the information about the symmetry potential at high density,
the beam energy required has to be of higher than several hundreds
of MeV per nucleon, but then the isospin effects on heavy ion
collisions would become negligible and are usually not considered.
However, in some special cases, like the one around the particle
emission threshold, it is found that the symmetry potential
affects the particle production
\cite{LiGQ96,LiGQ97,Cassing99,Fuchs03}, especially, the ratio
between the number of produced negatively and positively charged
particles may depend sensitively on the density dependence of the
symmetry potential.  B.A. Li \cite{LiBA02,LiBA03} found that, in
an isospin-dependent hadronic transport model, the $\pi ^{-}/\pi
^{+}$ ratio, as well as the neutron-proton differential collective
flow, were sensitive to the behavior of the nuclear symmetry
potential at high densities. More recently, within the framework
of relativistic Landau Vlasov transport method, Gaitanos et al.
\cite{Gaitanos04,Gaitanos03} found that when the beam energy was
higher than 2 AGeV the sensitivity of $\pi ^{-}/\pi ^{+}$ ratio to
the form of the symmetry potential is largely reduced at high
densities. In this paper, we attempt a further investigation of
the energy dependence of the sensitivity of the $\pi ^{-}/\pi
^{+}$ production ratio to the form of the symmetry potential in
the UrQMD model. Furthermore, we try to explore a new candidate in
terms of the $\Sigma^-/\Sigma^+$ ratio for probing the symmetry
potential in high density matter, which is in addition to the $\pi
^{-}/\pi ^{+}$ production ratio.

The production of $\Sigma^-$ and $\Sigma^+$ hyperons  is closely
related to the neutron-proton asymmetry of the projectile-target
system, which means that the symmetry potential of nuclear matter
will affect the $\Sigma^-/\Sigma^+$ ratio. Consequently, the
$\Sigma^-/\Sigma^+$ ratio in heavy ion collisions at high energies
may also carry the information about the density dependence of the
symmetry potential of nuclear matter. Furthermore, the
isospin-dependent part of the single-particle potential of
$\Sigma$ hyperon in a nuclear medium, the, so called, Lane
potential, depends on the isospin asymmetry of nuclear matter,
which might also influence the $\Sigma^-/\Sigma^+$ ratio. In turn,
this study of $\Sigma^-/\Sigma^+$ ratio might provide us with the
information about the isospin-dependent part of the $\Sigma$
hyperon single-particle potential.

In this paper, the study of $\Sigma$ hyperon production is limited
near its threshold (1.79 GeV for $\Sigma$ hyperon production in
free space through a process of the type $BB\rightarrow BKY$
reaction) in order to observe the effect of the symmetry potential
on the $\Sigma^-/\Sigma^+$ ratio. Specifically, we consider the
neutron-rich system $^{132}Sn+^{132}Sn$ and the nearly
isospin-symmetric system $^{112}Sn+^{112}Sn$ at three different
beam energies of $1.5A$ (the sub-threshold energy), $2.5A$, and
$3.5A$ GeV. For our calculations, the UrQMD model
\cite{Bass98,Bleicher99,Weber03,Reiter03}, version 1.3, is
adopted, using the 'hard' Skyrme-type EoS for reactions with beam
energies $E_b\le 4A$ GeV. We find that most of the UrQMD model
calculations can simultaneously reproduce many experimental
measurements, which offers a good platform for studying the
isospin effects at SIS energies.

The paper is arranged as follows. In section II, we give our
method of including the isospin-dependent part of the mean field
in the UrQMD transport model. In section III, the numerical
results of pion and $\Sigma$ hyperon production and the
corresponding ratios between the negatively and positively charged
particles are presented. Finally, in section IV, a brief summary
and discussion are given.

\section{The treatment of the isospin-dependent part of the mean field in the UrQMD model}
Since the isospin dependence of nucleon-nucleon interaction has
been introduced explicitly in the UrQMD model, in order to study
the isospin effects in heavy ion collisions, we have to introduce
the symmetry potential into the mean field. In UrQMD model, the
Skyrme and the Yukawa potentials are included in the iso-scalar
part of the mean field, where the Yukawa parameter is related to
Skyrme parameters. In infinite nuclear matter, the contribution of
Yukawa potential to the total energy acts like the two-body Skyrme
contribution \cite{Bass98}. The Coulomb potential is also
implemented explicitly. Similarly, a symmetry potential should
also be included in the mean-field part.

The bulk symmetry energy $e_{sym}$ in Eq. (\ref{ieos}) can be
expressed as
\begin{equation}
e_{sym}=S_0 F(u) \label{vsym},
\end{equation}
where $S_0$ is the symmetry energy at the normal density and
$u=\rho/\rho_0$ the reduced density. In this paper, we take $S_0=30$ MeV and, in
order to mimic the strong variation of the density dependence of
the symmetry energy at high densities, we adopt the form of $F(u)$ as
used in \cite{LiBA02}:
\begin{equation}
F(u)=\left\{
\begin{array}{l}
F_1=u^\gamma \hspace{1cm}  \gamma>0 \\
F_2=u\cdot\frac{a-u}{a-1} \hspace{1cm}a>1
\end{array}
\right. .\label{fu}
\end{equation}
Here $a$ is the reduced critical density. Note that when $u>a$,
the symmetry potential energy will be negative.  Similarly,
following \cite{LiBA02}, we take $\gamma=1$ for stiff symmetry
potential (stiff-sym.pot.), namely, the $F_1^{\gamma=1}$, and
$a=3$ for soft symmetry potential (soft-sym.pot.), namely, the
$F_2^{a=3}$. Apparently, $F_1^{\gamma=1}$ and $F_2^{a=3}$ give the
two extremes of the symmetry energy at high densities, as is
illustrated in the following for the chemical potential.

\begin{figure}
\includegraphics[angle=0,width=0.8\textwidth]{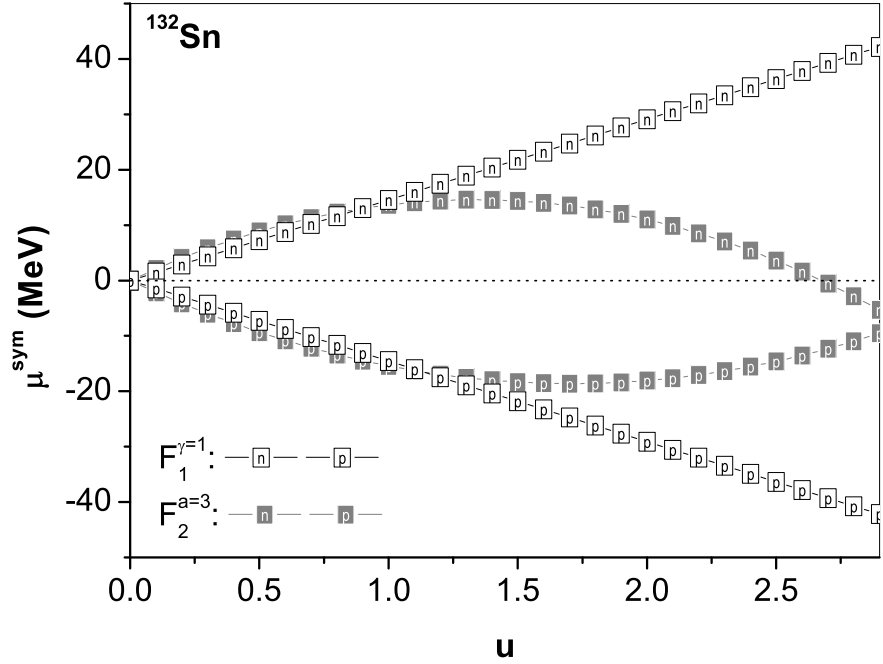}
\caption{The neutron and proton chemical potentials, contributed
only by the symmetry potential energy, calculated for $^{132}Sn$,
using both the stiff-sym.pot. $F_1^{\gamma=1}$ and soft-sym.pot.
$F_2^{a=3}$.  Here $\delta=(82-50)/132$.} \label{fig1}
\end{figure}

The neutron and proton chemical potentials $\mu_{n/p}^{sym}$, contributed only by the
symmetry potential energy, are shown in Fig.
\ref{fig1}.
We notice in Fig. \ref{fig1} that for $u<1$ the difference between
the neutron (and so also proton) chemical potentials calculated
with $F_1^{\gamma=1}$ and $F_2^{a=3}$ is small but becomes large
at high densities. For $u>2.6$, the $\mu_{n}^{sym}$ becomes
negative for $F_2^{a=3}$. Furthermore, the curves of
$\mu_{n}^{sym}$ for $F_1^{\gamma=1}$ and $F_2^{a=3}$ cross each
other at $u \sim 0.8$ (called, crossing point) and those of
$\mu_{p}^{sym}$ at $u \sim 1.1$. Apparently, such different
behaviors of the $\mu_{n/p}^{sym}$, calculated with
$F_1^{\gamma=1}$ and $F_2^{a=3}$, will strongly influence the
motion of protons and neutrons and so also the time evolution of
the proton and neutron density distributions, which will further
influence the ratios of produced particles with different charges.

For simplicity, the isospin-independent part of the mean field for
resonances and hyperons is taken here to be the same as that of
the nucleon. This simplification is quite adventurous, but we
think that it should not alter the final conclusion about the
influence of the symmetry potential on the ratios of emitted
negatively versus positively charged particles. The symmetry
potentials of the resonances [ N$^*$(1440) and $\Delta$(1232)] and
hyperons  [$\Lambda$ and $\Sigma$] are also introduced in the
calculations, in addition to that of the nucleon. The symmetry
potential for the resonances is obtained through the constants of
isospin coupling (the Clebsch-Gordan coefficients) in the process
of $\Delta(1232)$ [or $N^*(1440)$] $\leftrightarrow \pi N$.

For hyperons, the single-particle potential in a spin-saturated
nuclear matter can be expressed as
\begin{equation}
V_{\Sigma^{\pm}}=V_{0} \mp \frac{1}{2}V_{1}\delta , \label{vsigma}
\end{equation}
assuming charge independence of the baryon-baryon
interaction \cite{Lane,Dabrowski99}. Here $V_0$ and $V_1$ represent the
isospin-independent and -dependent parts. The $V_1$ is the Lane
potential, which is known to be important for the structure of the $\Sigma$
hyper-nuclear state\cite{Dov84,Har90}. This expression has the
same form as the single-particle potential of nucleons, up to the
first order in $\delta$. However, the value of $V_{1}$, even its
sign, is still a matter of argument: within the relativistic mean-field
theory (see, {\it e.g.}, \cite{Glen91,Scha96}), the symmetry
potential of the octet of baryons is described via the coupling
between the baryon and $\rho$ meson. For hyperons ($H$), the isospin
dependent part is determined by the coupling constant
$g_{H\rho}$ and their isospin. Since
$g_{H\rho}$ is taken to be $< g_{N\rho}$, as well as $> g_{N\rho}$ \cite{Glen91}, we simply take the symmetry
potential of $\Sigma^{\pm}$ hyperon to be proton-like and
neutron-like, according to Eq. (\ref{vsigma}). However, the symmetry
potential of the excited states of hyperon is not considered, for
lack of information.

Combining both the resonances and hyperons, we express the
symmetry potential in an unified form, which reads as
\begin{equation}
v_{sym}^{B}=\alpha v_{sym}^n+\beta v_{sym}^p ,\label{vx}
\end{equation}
where the values of $\alpha$ and $\beta$ for different baryons ($B$) are listed in
Table \ref{tab1}. From this table we can see that the symmetry
potentials of $\Delta^-$ and $\Sigma^{-}$ are neutron-like and
those of $\Delta^{++}$ and $\Sigma^{+}$ are proton-like. On the other hand,
the symmetry potentials of $\Delta^0$, $\Delta^+$ and
$N^*$(1440) are a mixture of the neutron and proton symmetry
potentials. Since the value of Lane potential is very uncertain, in order to make
comparisons, we have
also investigated the cases when the symmetry potential of
hyperons (and also of resonances) is switched off.

\begin{table}

\caption{The values of $\alpha$ and $\beta$ for isospin-dependent
potentials of different baryons.}

\begin{tabular}{|l|cc|l|cc|}
\hline\hline B  & $\alpha$ & $\beta$ & B  & $\alpha$ & $\beta$  \\
\hline
{N$^*$}$^0$(1440)&$1/3$ & $2/3$&$\Lambda$ & $1/2$ & $1/2$\\
{N$^*$}$^+$(1440)&$2/3$ &$1/3$&$\Delta^-$&$1$ & $0$\\
$\Sigma^-$ & $1$ & $0$& $\Delta^0$ &$2/3$ & $1/3$\\
$\Sigma^0$ & $1/2$ & $1/2$& $\Delta^+$&$1/3$ & $2/3$ \\
$\Sigma^+$ & $0$ & $1$&$\Delta^{++}$ &$0$ & $1$\\
\hline\hline

\end{tabular}
\label{tab1}
\end{table}

\section{Calculations and Results}
\subsection{Pion- and Hyperon-Production without symmetry potential}

In the UrQMD model, the total cross section depends on the
isospins of colliding particles, their flavour and the
center-of-mass energy. If high quality experimental data on the
respective cross sections (the neutron-proton, proton-proton
elastic scattering cross sections, {\it etc.}) exist, a
phenomenological fit to the respective data is adopted. If no data
are available, the Additive Quark Model (AQM) and the detailed
balance arguments are used. The details of the determination of
the elementary cross sections can be found in ref.
\cite{Bleicher99}.

\begin{figure}
\includegraphics[angle=0,width=0.7\textwidth]{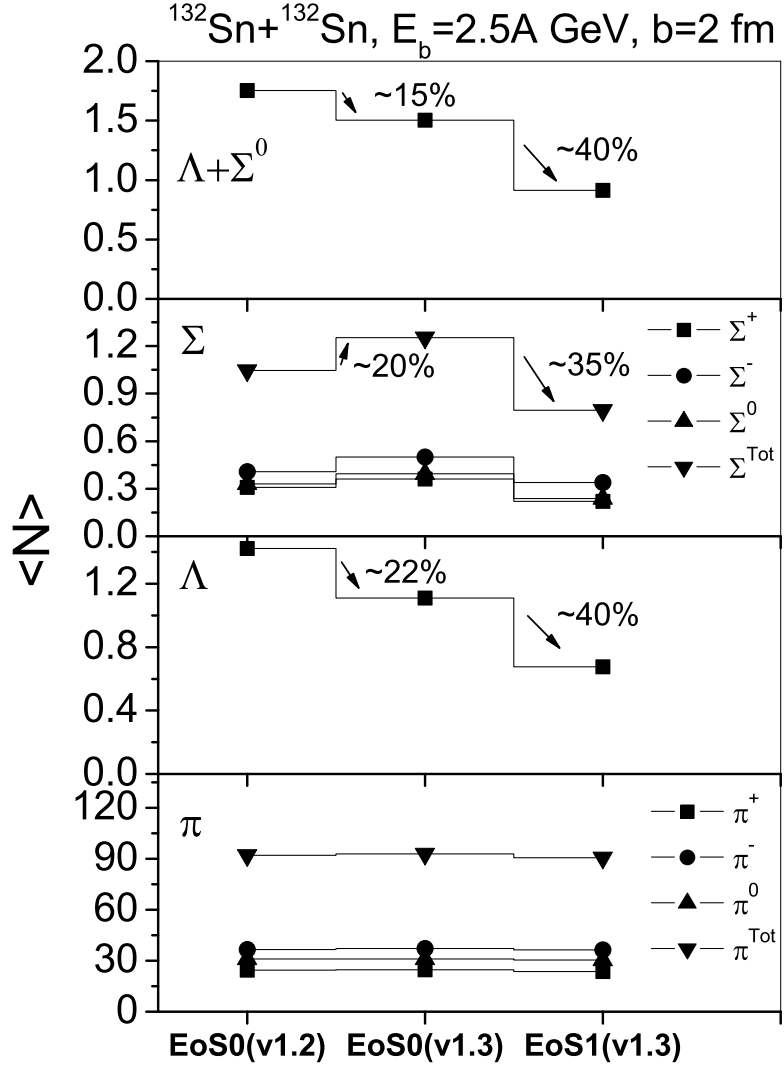}
\caption{The $\pi$, $\Lambda$, $\Sigma$, and ($\Lambda+\Sigma^0$)
yields with different conditions in the
$^{132}Sn+^{132}Sn$ reaction at $E_b=2.5A$ GeV and impact
parameter $b=2$ fm. EoS0 and EoS1 refer, respectively, to
without and with isoscalar part of the mean field, v1.2 and v1.3
being the version 1.2 and version 1.3 of UrQMD (see text).} \label{fig2}
\end{figure}

First of all, we check the pion and $\Sigma$ hyperon production
when the symmetry potential is not taken into account. In the
version 1.3 of UrQMD model, compared to the version 1.2, a channel
of $pp\rightarrow pK\Sigma$ is implemented for the improvement of
the kaon production at low energies. The introduction of this
channel may also influence the yields of hyperons. In Fig.
\ref{fig2}, we show the yields of pions, $\Lambda$, and $\Sigma$
hyperons  in the reaction $^{132}Sn+^{132}Sn$ at $E_b=2.5A$ GeV
and $b=2$ fm, calculated by using both the versions 1.2 and 1.3 of
UrQMD, and with and without iso-scalar part of the mean field. The
EoS0(v1.2) and EoS0(v1.3) denote the results of the versions 1.2
and 1.3 without the mean field, and the EoS1(v1.3) denotes the
results of the version 1.3 with a mean field of hard Skyrme-type
force. The freeze-out time is taken to be $30 fm/c$ and for each
case we calculate tens of thousand events for good statistics.  In
Fig. \ref{fig2} (lower panel) we find that the effect of
introducing the process of $pp\rightarrow pK\Sigma$ and the mean
field on $\pi$ production is almost negligible, but its effect on
the hyperon production is rather large (see upper panels in Fig.
\ref{fig2}). The yield of $\Sigma$ hyperon is enhanced by about
$20\%$ whereas the yield of $\Lambda$ hyperon is suppressed by
about $22\%$ with the version 1.3, which together suppresses the
$\Lambda+\Sigma^0$ yields by about $15\%$. When the mean field is
switched on, the suppression effect on the yields of hyperons is
enhanced by about $35-40\%$, the same as in \cite{Reiter03}. In
Refs. \cite{Bass98,Bleicher99,Reiter03}, it was pointed out that
the yields of hyperons within the cascade model were always
overestimated. This means that our suppressed yields of the
hyperon production, with mean-field correction, could be in the
right direction of the experiments at these energies.

\begin{figure}
\includegraphics[angle=0,width=0.8\textwidth]{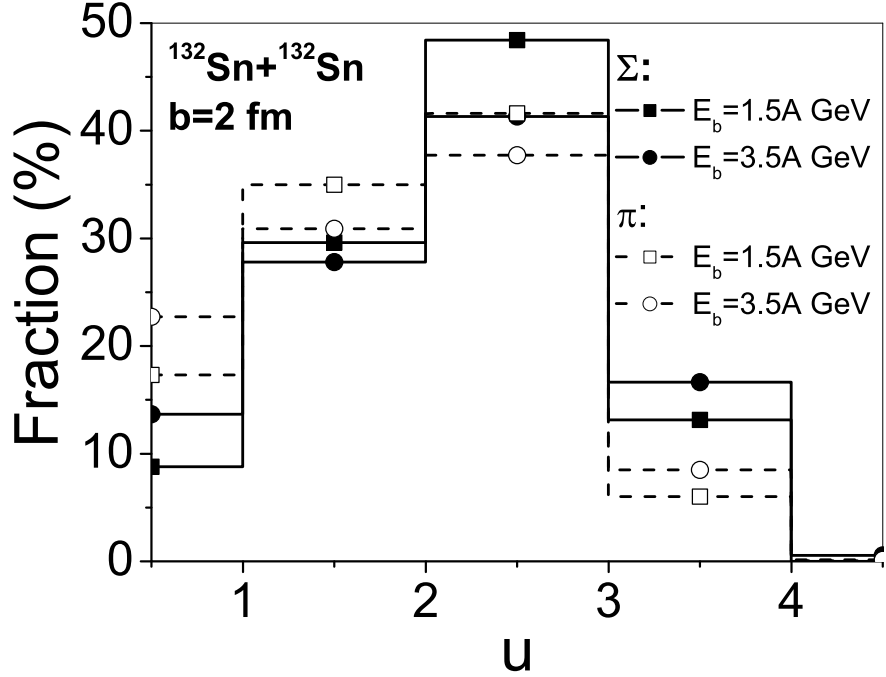}
 \caption{The production fractions of $\pi$ and $\Sigma$
at different densities for $^{132}Sn+^{132}Sn$ reactions at $E_b=1.5A$ and $3.5A$ GeV and $b=2$
fm.} \label{fig3}
\end{figure}

In Fig. \ref{fig3}, we show the fractions of the yields of pions
and $\Sigma$ hyperons produced in different baryon density
regions, for cases of $E_b=1.5A$ and $3.5A$ GeV. Here the
percentage of the contributions in the individual density regions
to the total yield is plotted. In general, pions and $\Sigma$
hyperons are mainly produced in the high density region ($u>1$). Our
calculations at both the energies show that more than $75\%$ of
pions and $\Sigma$ hyperons are produced in the region of $u>1$.
Among them, more than $50\%$ is produced in the region of $u=1-3$,
and the remaining is produced in the region of $u>3$. This happens
because the phase-space available at such higher densities is
quite restricted. Furthermore, with the energy increasing from
$1.5A$ GeV to $3.5A$ GeV, the fractions of the yields of $\pi$
(and $\Sigma$) produced in the density region of $u=1-3$ decrease
while those at other densities increase. Comparing the pion and
$\Sigma$ production, we find that at both energies the fraction of
pions produced is larger than that of $\Sigma$ hyperons for
densities $u<2$, while the situation is just opposite for
densities $u>2$.

\begin{figure}
\includegraphics[angle=0,width=1.0\textwidth]{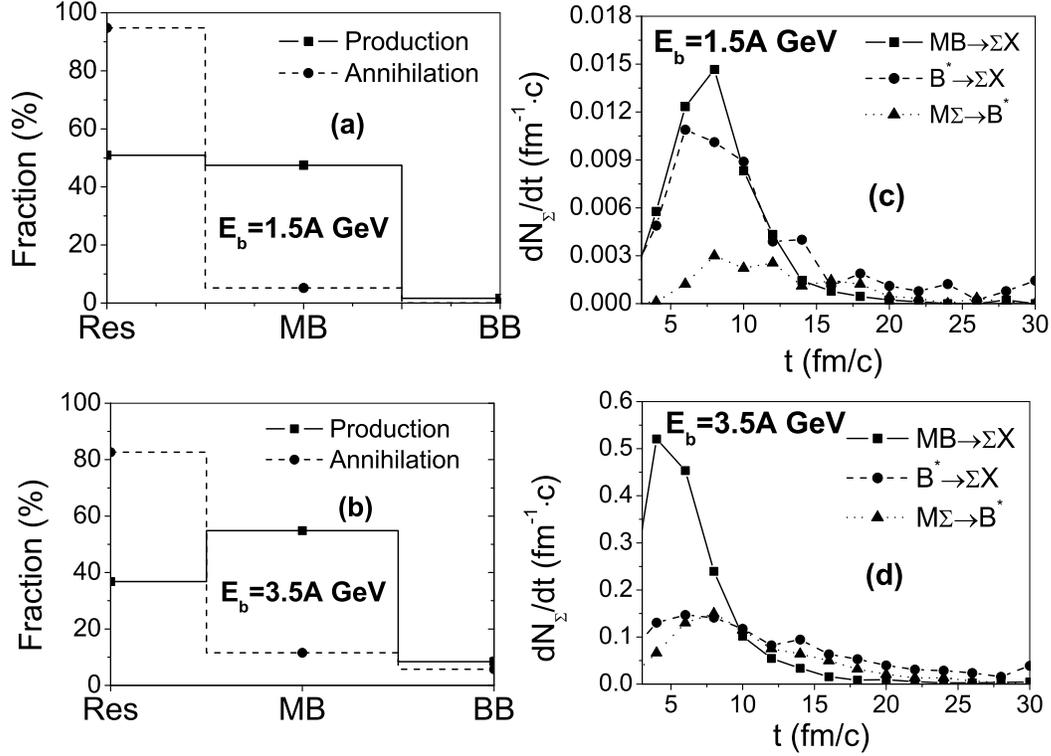}
\caption{The contributed ratios of each channel (BB, MB or Res) for
$\Sigma$ production or annihilation plotted at (a) $E_b=1.5A$ GeV and (b) $3.5A$ GeV. The time
evolution of d$N_{\Sigma}$/dt of  several important $\Sigma$
production and annihilation processes is shown at (c) $E_b=1.5A$ GeV and (d) $E_b=3.5A$ GeV.}
\label{fig4}
\end{figure}

Figs. \ref{fig4} (a) and (b) show the percentage of the contributions from the
relevant Baryon-Baryon (B-B), Meson-Baryon (M-B) inelastic
scattering, and the resonance (Res) decay processes to the total
$\Sigma$ hyperon production and annihilation at (a) $E_b=1.5A$ GeV
and (b) $E_b=3.5A$ GeV. It should be pointed out that in these results,
the contributions from the $B\Sigma\rightarrow \Sigma X$ in
$BB\rightarrow \Sigma X$ and the $M\Sigma\rightarrow \Sigma X$ in
$MB\rightarrow \Sigma X$ are added to both the $\Sigma$
hyperon production and annihilation processes. Here, and below, "X"
represents the products (including multi-hadrons) other than
$\Sigma$. From these two plots, we find that the most important
channels for $\Sigma$ production and annihilation are
$MB\rightarrow \Sigma X$, and the Res ($B^*\rightarrow \Sigma X$, and $\Sigma
M\rightarrow B^*$) at the two energies studied.

Figs. \ref{fig4} (c) and (d) show the time evolution of the
average number of $\Sigma$ produced and annihilated per unit time,
$dN_{\Sigma}/dt$, through the processes of $MB\rightarrow \Sigma
X$, $B^*\rightarrow \Sigma X$, and $M\Sigma \rightarrow B^*$ at
$E_b=1.5A$ GeV and $E_b=3.5A$ GeV, respectively. Here, the unit
time is taken to be $2fm/c$. For $\Sigma$ production, at the early
stage, the channel $MB\rightarrow\Sigma X$ is the most important
one. The $dN_{\Sigma}/dt$ for $MB\rightarrow \Sigma X$,
particularly at $E_b=3.5A$ GeV, is very much pronounced at the
early reaction time, but then it decreases, even faster than for
the case of $1.5A$ GeV. The $dN_{\Sigma}/dt$ for $MB\rightarrow
\Sigma X$ is reduced to $\sim 20\%$ of its highest value at
$t=10fm/c$ for the case of $E_b=3.5A$ GeV but at $t=12fm/c$ for
the case of $E_b=1.5A$ GeV.  At the late reaction stage, $\Sigma$
is mainly produced from the decay of baryon resonances (i.e.
$\Sigma^*$ or $\Lambda^*$), which then continues for a longer
time. Concerning the $dN_{\Sigma}/dt$-value for the $\Sigma$
annihilating channel of $M\Sigma\rightarrow B^*$, it is much
smaller compared to the case of the reverse process of
$B^*\rightarrow \Sigma X$ at $E_b=1.5A$ GeV, while it remains
comparable at $E_b=3.5A$ GeV. The reason for this behavior is that
a larger number of $\Sigma$'s are produced through $MB\rightarrow
\Sigma X$ at $E_b=3.5A$ GeV than at $E_b=1.5A$ GeV, which leads to
a stronger annihilation of $\Sigma$'s at $E_b=3.5A$ GeV. Thus,
from Figs. \ref{fig4} (c) and (d) we may conclude that $\Sigma$
hyperons are mainly produced during $t < 10$ fm/c for the case of
$E_b=3.5A$ GeV but continues up to $t \sim 12 $fm/c for the case of
$E_b=1.5A$ GeV.

\begin{figure}
\includegraphics[angle=0,width=0.8\textwidth]{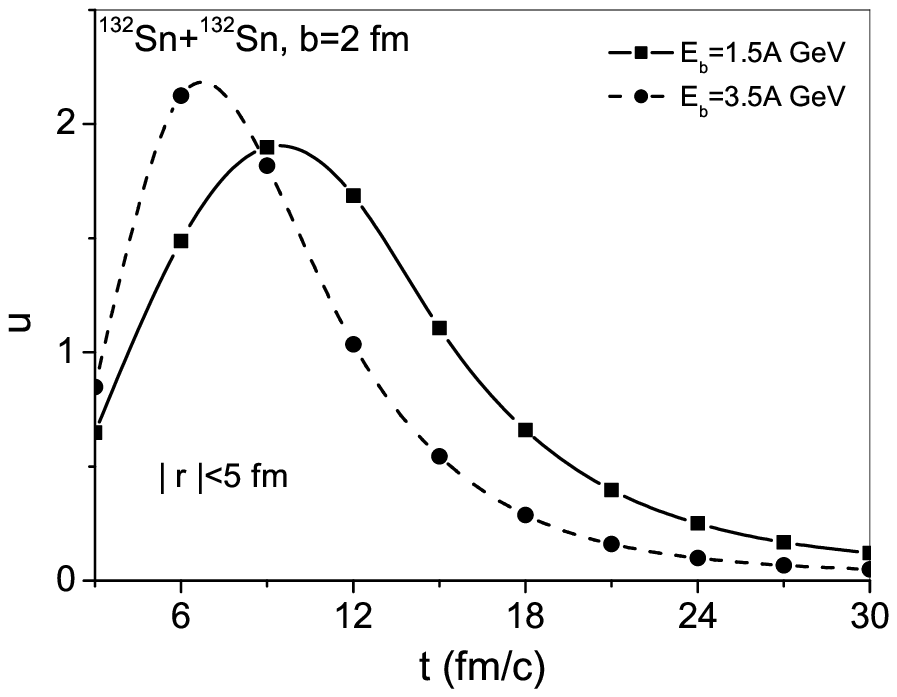}
\caption{The time evolution of the average density in central
collision zone at $E_b=1.5A$ and $3.5A$ GeV.}
\label{fig5}
\end{figure}

Fig. \ref{fig5} shows the time evolution of the average density in
the central reaction zone of $|r|<5$ fm at $E_b=1.5A$ and $3.5A$
GeV. One sees that the average density in the region of $|r|<5$ fm
reduces to the normal density at $t\sim 12$ fm/c for the case of
$E_b=3.5A$ GeV, but for $E_b=1.5A$ GeV at $t\sim 16fm/c$. Compared to the case when $dN_{\Sigma}/dt$
for the process $MB\rightarrow \Sigma X$ reduces to $\sim 20\%$ of
its highest value,  this result at high energy is similar but is somewhat longer at the lower energy.
In other word, Figs. \ref{fig4} and \ref{fig5} show that, compared to the time when the system stays at
high densities ($u>1$), the time when most of the $\Sigma$
hyperons are produced is similar at $E_b=3.5A$ GeV but is somewhat
{\it shorter} at $E_b=1.5A$ GeV.

\begin{figure}
\includegraphics[angle=0,width=0.9\textwidth]{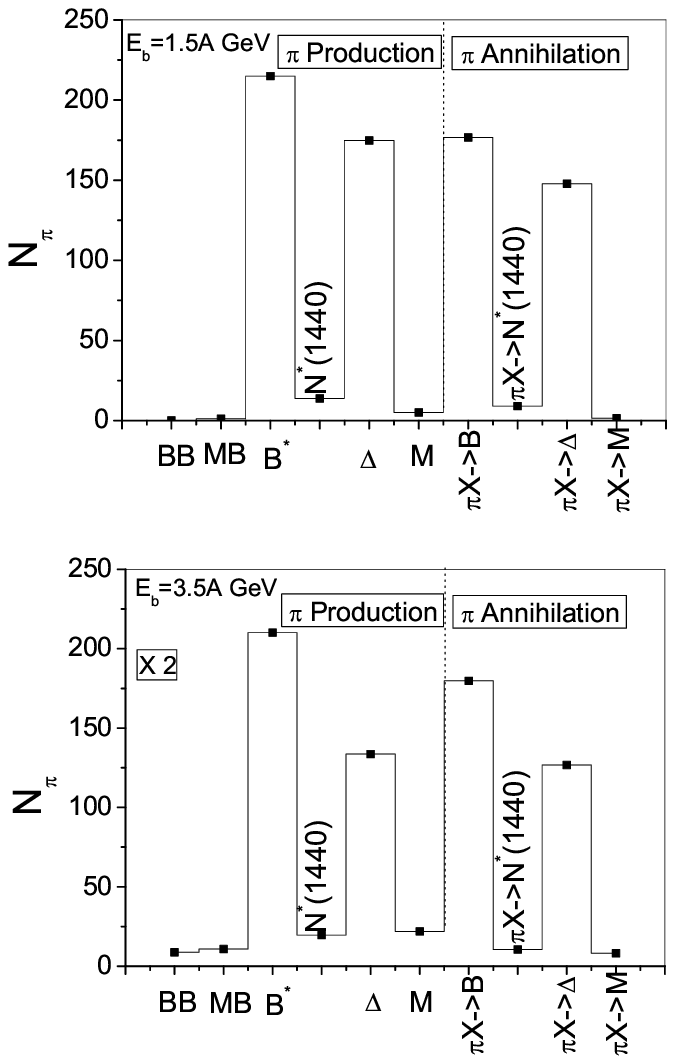}
 \caption{The average number of $\pi$ produced
and annihilated through different channels for $^{132}Sn +
^{132}Sn$ at $E_b=1.5A$ GeV (upper plot) and $3.5A$ GeV (lower
plot). For a comparison between the two plots, the
number in lower plot is drawn as half of the real quantity. }
\label{fig6}
\end{figure}

Similar to Figs. \ref{fig4} (a) and (b), Fig. \ref{fig6} shows the
accumulated number of $\pi$'s produced and annihilated via the
different processes. Unlike $\Sigma$, pion is mainly produced
via the baryon decay, where the most important process is
$\Delta$ decay. The contribution from other channels, such as
$BB$, $MB$, $M$ decay, as well as $N^*(1440)$ decay, becomes
visible only for the case of $3.5A$ GeV. The obvious difference
between the two cases of $E_b=1.5A$ GeV and $3.5A$ GeV is that the
{\it fraction} of $\pi$ produced via $\Delta$ decay is
much smaller for the higher energy case. A similar scenario occurs
for $\pi$ annihilation.

\subsection{The effect of the symmetry potential on $\pi^-/\pi^+$ and $\Sigma^-/\Sigma^+$ ratios}
\begin{figure}
\includegraphics[angle=0,width=0.8\textwidth]{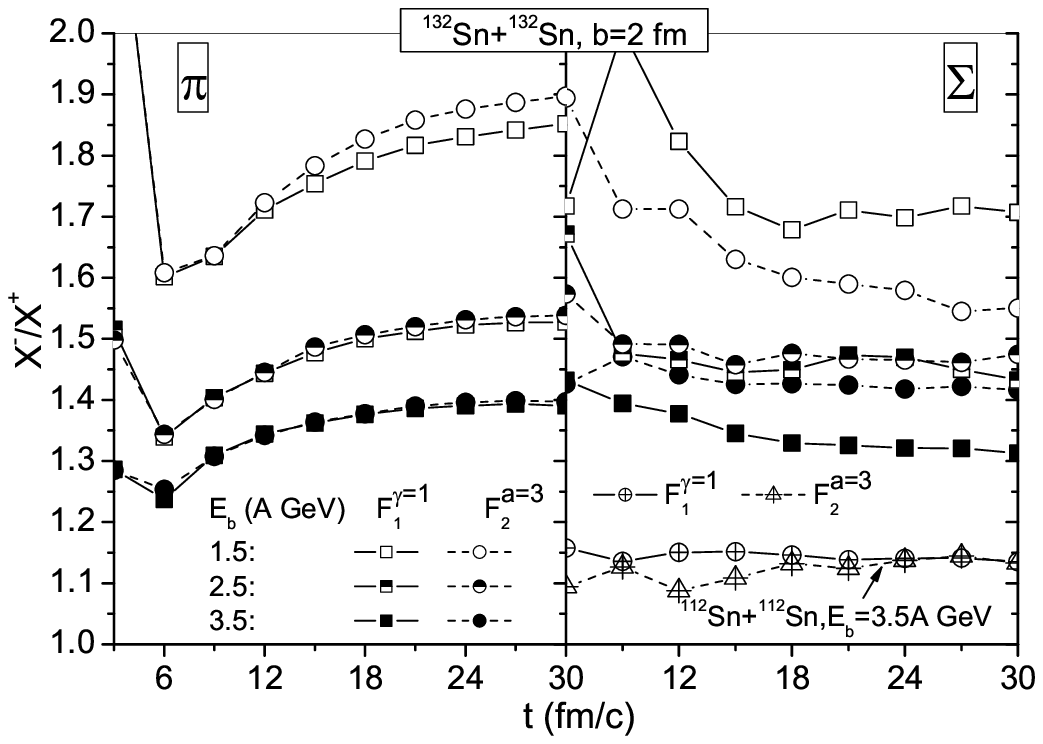}
\caption{The ratios $\pi^-/\pi^+$ (left) and $\Sigma^-/\Sigma^+$
(right) for the collisions $^{132}Sn+^{132}Sn$ ($E_b=1.5A$,
$2.5A$, and $3.5A$ GeV; $b=2$ fm) and $^{112}Sn+^{112}Sn$
($E_b=3.5A$ GeV; $b=2$ fm), calculated with the different symmetry
potentials $F_1^{\gamma=1}$ and $F_2^{a=3}$.} \label{fig7}
\end{figure}

In this section, we  mainly explore the effect of the symmetry
potential on the ratios between the yields of negatively and
positively charged pions and $\Sigma$ hyperons. In Fig.
\ref{fig7}, we show the time evolution of the $\pi^-/\pi^+$ ratios
(left-hand side) and the $\Sigma^-/\Sigma^+$ ratios (right-hand
side) calculated with $F_1^{\gamma=1}$  and $F_2^{a=3}$ for the
reaction $^{132}Sn+^{132}Sn$ at $E_b=1.5A$, $2.5A$, $3.5A$ GeV and
$b=2$ fm, and $^{112}Sn+^{112}Sn$ at $E_b=3.5A$ GeV and $b=2$ fm.
The symmetry potentials of all the particles mentioned in Sec. II
are considered. Here, and below, the $\Sigma$ resonances are also
included in order to improve the statistics of calculated
quantities at the early stage of the reaction. Also, at the
freeze-out time, all the unstable particles are considered to
decay. For the pions, one can see from the left plot in Fig.
\ref{fig7} that at $E_b=1.5A$ GeV the $\pi^-/\pi^+$ ratio calculated with the soft
symmetry potential ($F_2^{a=3}$) is enhanced compared with the
stiff one ($F_1^{\gamma=1}$), which is the same
as was found in \cite{LiBA02}. This happens because, at this
energy, pions are mainly produced by $\Delta$ decay (refer to Fig.
\ref{fig6}). Furthermore, we also find that the
difference in the ratios of $\pi^-/\pi^+$ calculated with
different symmetry potentials $F_1^{\gamma=1}$ and $F_2^{a=3}$ is
reduced with the increase of the beam energy, like in Ref. \cite{Gaitanos03}. The reason for this
insensitivity at high energies is that more $\pi$ production and
annihilation channels are involved at high energies and then the
$\pi$ production and annihilation via $\Delta$-decay also become
less important. In \cite{Gaitanos03}, it is mentioned that the
recent FOPI data show this above noted tendency. Thus, we could take the ratio
$\pi^-/\pi^+$ at high energies (as is the case with $E_b=3.5A$ GeV
studied in this work) to have become insensitive to the symmetry
potential.

Next, for $\Sigma$ hyperons, from Fig. \ref{fig7} (right plot),
firstly, one sees that the $\Sigma^-/\Sigma^+$ ratio is sensitive
to the density dependence of the symmetry potential for
neutron-rich $^{132}Sn+^{132}Sn$ collisions, but insensitive to
that for the nearly symmetric $^{112}Sn+^{112}Sn$ collisions. For
$^{132}Sn+^{132}Sn$ at $E_b=1.5A$ GeV, the $\Sigma^-/\Sigma^+$
ratio calculated with the stiff symmetry potential is higher than the
one with the soft symmetry potential. As the beam energy
increases, the $\Sigma^-/\Sigma^+$ ratio falls and the difference
between the $\Sigma^-/\Sigma^+$ ratios calculated with
$F_1^{\gamma=1}$ and $F_2^{a=3}$ reduces strongly. As the beam
energy increases further, at $E_b=3.5A$ GeV the
$\Sigma^-/\Sigma^+$ ratio falls further but the difference between
the $\Sigma^-/\Sigma^+$ ratios calculated with $F_1^{\gamma=1}$
and $F_2^{a=3}$ appears again, the $\Sigma^-/\Sigma^+$
ratio with soft symmetry potential now becoming higher than that with
the stiff one.

\begin{figure}
\includegraphics[angle=0,width=0.8\textwidth]{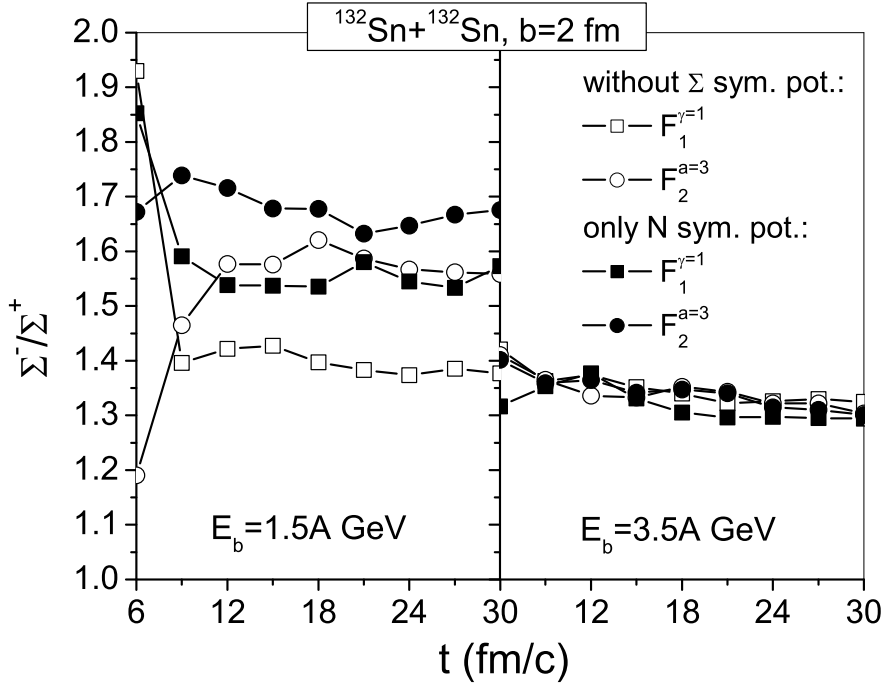}
 \caption{The ratios $\Sigma^-/\Sigma^+$
for $F_1^{\gamma=1}$ and $F_2^{a=3}$, with the symmetry potentials
of different baryons switched on or off.} \label{fig8}
\end{figure}

In sequel, it might be interesting to investigate why the behavior
of $\Sigma^-/\Sigma^+$ and $\pi^-/\pi^+$ ratios is so
different, as far as the influence of the density dependence of
the symmetry potential is concerned. One basic difference is that,
like nucleons, $\Sigma^{\pm}$ hyperons are under the influence of
the mean field produced by the surrounding nucleons, as soon as
they are produced. The symmetry potential of hyperons also
play an important dynamic role and results in a strong effect on
the ratio of the negatively to positively charged $\Sigma$
hyperons. Thus, we further investigate the $\Sigma^-/\Sigma^+$ and
$\pi^-/\pi^+$ production ratios when the symmetry potential of
$\Sigma$ hyperons and resonances, except nucleons, is switched
off. A very small difference for the $\pi^-/\pi^+$ ratio, but a
large difference for $\Sigma^-/\Sigma^+$ ratio is found with the
switching on and off of the symmetry potential of $\Sigma$ and
resonances. Our results for $\Sigma^-/\Sigma^+$ ratio are plotted
in Fig. \ref{fig8}. Two cases
are demonstrated: 1) only the symmetry potential of nucleons is
considered; and 2) only the $\Sigma$ symmetry potential is not
considered.

From Fig. \ref{fig8} one sees that the $\Sigma^-/\Sigma^+$
ratio at $E_b=1.5A$ GeV with soft symmetry potential is higher
than with the stiff symmetry potential, no matter whether the
symmetry potential of $N^*(1440)$, $\Delta$, and $\Lambda$ is
introduced or not. As the incident energy is increased to $3.5A$
GeV, the sensitivity to the density dependence of the symmetry
potential is almost lost, like for the case of
$\pi^-/\pi^+$ ratio (not shown here). This means that the
different behavior of the $\Sigma^-/\Sigma^+$ ratio, with respect
to the $\pi^-/\pi^+$ ratio in Figs. \ref{fig7}, is due to the introduction of the
symmetry potential of $\Sigma$ hyperon. This can be understood as
follows: In Fig. \ref{fig4}, it was shown that one of the most
important channel for $\Sigma$ production is $MB\rightarrow \Sigma
X$, where the most important channel is $\pi N\rightarrow \Sigma
X$. There are primarily 6 processes for $\pi N\rightarrow \Sigma
X$, namely, 1) $\pi^{+} n\rightarrow \Sigma^{0}K^{+}$, 2) $\pi^{+}
n\rightarrow \Sigma^{+}K^{0}$, 3) $\pi^{-} n\rightarrow
\Sigma^{-}K^{0}$, 4) $\pi^{-} p\rightarrow \Sigma^{-}K^{+}$, 5)
$\pi^{-} p\rightarrow \Sigma^{0}K^{0}$, 6) $\pi^{+} p\rightarrow
\Sigma^{+}K^{+}$. The 1) and 5) are not relevant here; 2) and 6)
are relevant to the $\Sigma^+$ production and 3) and 4) are
relevant to the $\Sigma^-$ production. Thus, one can draw the
conclusion that the $\Sigma^-/\Sigma^+$ ratio should be
proportional to the $\pi^-/\pi^+$ ratio, and hence, the
$\Sigma^-/\Sigma^+$ ratio has the same behavior as the
$\pi^-/\pi^+$ ratio when the symmetry potential of $\Sigma$ is not
taken into account.

\section{Summary and Discussion}

In summary, based on the UrQMD model (version 1.3), we have
investigated the influence of the symmetry potential on the ratios
between the negatively and positively charged pions and $\Sigma$
hyperons, in the nearly central collisions $^{132}Sn+^{132}Sn$ and
$^{112}Sn+^{112}Sn$ at $1.5A$, $2.5A$ and $3.5A$ GeV energies. In
order to find sensitive probes to the behavior of the symmetry
potential at high-density nuclear matter, two different forms of
the density dependence of symmetry potential in the mean field
are considered. The obvious dynamical effect of the symmetry
potential is found on the neutron-rich reaction
$^{132}Sn+^{132}Sn$ and not on the nearly isospin-symmetric
reaction $^{112}Sn+^{112}Sn$. The effect of the symmetry potential
on the $\pi^-/\pi^+$ ratio in
 $^{132}Sn+^{132}Sn$ at $E_b=1.5A$ GeV is similar to that found in
 \cite{LiBA02,Gaitanos03}, namely, the $\pi^-/\pi^+$ ratio
calculated with the soft symmetry potential is higher than that
with the stiff one, but at higher energies, like $E_b=3.5A$ GeV,
it disappears. This is explained as follows: at $E_b=1.5A$ GeV the
most important channel for the production of pions is $\Delta$
decay, while at $E_b=3.5A$ GeV other channels also play important
role and the contribution from $\Delta$ decay is largely
reduced.

The situation about the effect of the symmetry potential on
$\Sigma^-/\Sigma^+$ is more complicated because $\Sigma$ hyperon
itself also experiences a mean field of nuclear medium as soon as
it is produced. When the symmetry potential of $\Sigma$ hyperons
is not taken into account, a behavior similar to that of
$\pi^-/\pi^+$ ratio is obtained, i.e., the $\Sigma^-/\Sigma^+$
ratio calculated with the soft symmetry potential is higher than
that with the stiff one at $E_b=1.5A$ GeV and the sensitivity to
the symmetry potential disappears at $E_b=3.5A$ GeV.

As soon as the symmetry potential of $\Sigma$ is introduced, the
motions of $\Sigma^{-}$ and $\Sigma^{+}$ are also governed by the
symmetry potential, in addition to the iso-scalar part of the
single-particle potential. The density dependence of
$\mu^{sym}$ (see Fig. \ref{fig1}, where $\mu^{sym}$ of
$\Sigma^{-}$ is similar to that of neutron and that of $\Sigma^{+}$ is
similar to that of proton) drives $\Sigma^{-}$
($\Sigma^{+}$) to high (low) density area for soft symmetry
potential and to low (high) density area for stiff symmetry
potential, when the density is higher than the crossing density,
and vice versa when the density is lower than the crossing density,
though the effect is then much smaller. Simultaneously, $\Sigma^{-}$
($\Sigma^{+}$) with soft symmetry potential possesses lower
(higher) symmetry potential energy than that with stiff symmetry
potential, when density is higher than the crossing density, and
vice versa when density is lower than the crossing density. For
the case of $E_b=1.5A$ GeV, the time duration at high densities
($u > 1$) is much longer than the time when the most of $\Sigma$
hyperons are produced. Therefore, the situation in this case is
mostly like that of the case when the density is higher than the crossing
density, i.e., $\Sigma^{-}$ hyperons move to the high density area
and $\Sigma^{+}$ hyperons move to the low density area for the
soft symmetry potential. Thus, with the soft symmetry
potential, the annihilation of $\Sigma^{-}$ is enhanced and that
of $\Sigma^{+}$ is reduced, compared with those with stiff symmetry
potential. Furthermore, the single-particle energy of $\Sigma^{-}$
with the stiff symmetry potential is higher than that with the
soft symmetry potential,  which leads to more $\Sigma^{-}$
hyperons emitted for the stiff symmetry potential case than for
the soft one. Finally, the $\Sigma^-/\Sigma^+$ ratio with the
stiff symmetry potential may exceed the one with that of the soft
symmetry potential. On the other hand, for the case of $E_b=3.5A$ GeV, the
duration time at high density is much shorter, as is also the case for the
$\Sigma$ hyperons produced at high densities. Also, the nuclear density
reduces quickly and the most of $\Sigma$ hyperons after their
production will experience the situation with the density being
lower than the crossing density. Consequently, the
$\Sigma^-/\Sigma^+$ ratio with the soft symmetry potential may
exceed as compared to that with the stiff symmetry potential. This kind of energy
dependence of the behavior of the $\Sigma^-/\Sigma^+$ ratio with
respect to the symmetry potential is completely due to the
dynamical effect of the symmetry potential of $\Sigma$ in nuclear
medium, which allows us to understand why the behavior of the
$\Sigma^-/\Sigma^+$ ratio is different from the $\pi^-/\pi^+$
ratio. However, there exists a large uncertainty about the
single-particle potential of $\Sigma$, especially the symmetry
potential, so that the results presented here may not be accurate
quantitatively. However, all the features about the energy dependence
of the relative values of the $\Sigma^-/\Sigma^+$ ratios,
corresponding to the different forms of the density dependence of
the symmetry potential, should not change. These features are also
useful for us to extract the information about the symmetry
potential of $\Sigma$ hyperon.

\section*{Acknowledgments}
This work is supported by the National Natural Science Foundation of
China under Grant Nos. 10175093 and 10235030, Major State
Basic Research Development Program under Contract No. G20000774,
the Knowledge Innovation Project of the Chinese Academy of
Sciences under Grant No. KJCX2-SW-N02, the CASK.C. Wong
Post-doctors Research Award Fund, and the Alexander von Humboldt
Foundation, Germany.

\end{document}